\documentclass[12pt,preprint]{aastex}

\shorttitle{SN2010U - a nova}
\shortauthors{Humphreys et al.}

\begin{document}


\title{SN2010U -- a Luminous Nova in NGC 4214\altaffilmark{1}} 


\author{Roberta M. Humphreys\altaffilmark{2}, 
Jos\'e L. Prieto,\altaffilmark{3,7}
Philip Rosenfield,\altaffilmark{4}   
L. Andrew Helton,\altaffilmark{2} 
Christopher S. Kochanek,\altaffilmark{5} 
K. Z. Stanek,\altaffilmark{5}
Rubab Khan,\altaffilmark{5} 
Dorota Szczygiel,\altaffilmark{5} 
Karen Mogren,\altaffilmark{5}
Robert A. Fesen,\altaffilmark{6} 
Dan Milisavljevic,\altaffilmark{6} 
Benjamin Williams,\altaffilmark{4} 
Jeremiah Murphy,\altaffilmark{4} 
Julianne Dalcanton,\altaffilmark{4} 
Karoline Gilbert,\altaffilmark{4} 
}

\altaffiltext{1}
{Some of the data presented in this paper were obtained from the Multimission Archive at the Space Telescope Science Institute (MAST). STScI is operated by the Association of Universities for Research in Astronomy, Inc., under NASA contract NAS5-26555. Support for MAST for non-HST data is provided by the NASA Office of Space Science via grant NNX09AF08G and by other grants and contracts.} 
\altaffiltext{2}
{Astronomy Department, University of Minnesota, Minneapolis, MN 55455}
\altaffiltext{3}
{Carnegie Observatories, Pasadena, CA 91101}
\altaffiltext{4}
{Astronomy Department, University of Washington, Seattle, WA 98195}
\altaffiltext{5}
{Astronomy Department, The Ohio State University, Columbus, OH 43210}
\altaffiltext{6}
{Department of Physics \& Astronomy, Dartmouth College, Hanover, NH 03755} 
\altaffiltext{7}
{Hubble and Carnegie-Princeton Fellow}

\email{roberta@umn.edu}

\begin{abstract}
The luminosity, light curve, post--maximum spectrum, and lack of a progenitor
on deep pre-outburst images suggest that SN 2010U was a luminous, fast nova.
Its outburst magnitude is consistent with that  for a fast nova using the Maximum Magnitude-Rate of Decline relationship for classical novae.
\end{abstract}



\section{Introduction}

Episodic high mass loss events dominate the final stages of massive star 
evolution. The evidence for these events is observed across the upper HR 
diagram ranging from giant eruptions like $\eta$ Car, and the ``LBV 
nebulae'' associated with many of the Luminous Blue Variables to the 
cool hypergiants and red supergiants like IRC~+10420 and VY CMa with 
their complex circumstellar ejecta.  Unfortunately, the observational 
record is sparse because these stars are rare and their importance has 
only been fully recognized in recent years.

Modern supernova surveys are 
finding  a growing assortment of similar and related objects. 
These ``impostors'' are under-luminous, have spectra with narrow emission 
lines and much slower ejection velocities. Some of these objects may be undergoing 
`giant' eruptions possibly similar to $\eta$ Car but with much shorter 
durations (e.g., SN 2009ip, \citet{Smi2010}). Others appear to be generically related 
to the normal LBV/S Dor variables like SN 2002kg (Var 37 in NGC 2403; 
\citet{KW05,vandyk06,Maund06}). 
A third subgroup including SN 2008S and the NGC 300 optical transient 
 were heavily obscured prior to the outburst \citep{JPetal08, JP08}; their progenitors  may have 
recently evolved from a high mass losing stage such as an AGB star or red 
supergiant \citep{Smi09,Bond09,Berg09,Bott09,Thomp09,JPetal09}. We used to think that ``eruptions''  in which the star increased 
its total luminosity were primarily associated with the most massive stars, 
upwards of 40 --50 M$\odot$, but the latter pair apparently originated 
from 10 -- 20 M$\odot$ stars (references above and \citet{Gogarten}). 
It is thus increasingly apparent that these eruptions or outbursts may not all 
be the same phenomenon; the objects  represent a range of stellar masses and 
may originate from stars in different evolutionary states. 

Although discovery of these transient non-terminal eruptions are becoming more 
common, our information  about them is still very incomplete. 
The most recent addition, SN2010U in NGC 4214,  was first reported in eruption on 
5 February, 2010 (S. Nakano, CBET 2161) at apparent magnitude $\sim$ 16 on 
unfiltered CCD frames.  An early spectrum (Marion, Vinko, \& Wheeler, CBET 2163) with strong, narrow emission lines  of hydrogen, Ca II
and Na I with P Cyg absorptions, quickly showed that it was not a true supernova, but possibly an 
eruptive variable or transient. 

In this {\it Letter} we  present pre and post-eruption observations 
of SN 2010U and a analysis of the associated stellar population in NGC 4214.
Our discussion of its light curve, spectrum,  and  limitations on  the mass
and luminosity of its likely progenitor suggest that SN2010U is a very luminous nova, not the eruption of a massive star.

\section{Pre- and Post--Eruption Observations}

The discovery photometry for SN 2010U reported in CBET 2161 listed magnitudes of 
16.0 (Feb 5.6), 16.3 (Feb 5.7), 15.9 (Feb 6.5) amd 16.3 (Feb 6.6) on unfiltered
CCD frames from different observers. However  a fainter  R magnitude of 17.3 was 
reported by Brimacombe also on Feb. 6, 2010\footnote{http://www.rochesterastronomy.org/sn2010/index.html.}.   
This difference of more than a magnitude in such a short time seemed too large to be realistic. At our request,  the observers kindly provided their unfiltered CCD images.
We determined the magnitude of SN 2010U relative to three SDSS stars in the field 
with gri photometry converted to broadband R magnitudes. The photometry was 
done on galaxy-subtracted images, using a SDSS r-band image for the template galaxy  \citep{WF09}. The results yield an 
R band magnitude near 17~mag $\pm$ 0.2 -- 0.3~mag. 
We also obtained post-eruption photometry of SN 2010U with Retrocam 
on the MDM\footnote{The MDM Observatory is owned and operated by a consortium of five universities: the University of Michigan, Dartmouth College, the Ohio State University, Columbia University, and Ohio University.} Hiltner 2.4m telescope on Kitt Peak and the  Large Binocular Camera  \citep{Giallongo08} on the 
LBT\footnote{The Large Binocular Telescope is an international collaboration among institutions
in the United States, Italy and Germany. The LBT Corporation partners are:
the University of Arizona on behalf of the Arizona university system;
the Istituto Nazionale di Astrofisica, Italy; the
LBT Beteiligungsgesellschaft, Germany, representing the Max Planck Society, the Astrophysical Institute Potsdam, and Heidelberg University;
the Ohio State University; and
the Research Corporation, on behalf of the University of Minnesota, University of University of Notre Dame, and University of Virginia.}
on Mt. Graham. The journal of observations and measured magnitudes for SN 2010U
are included in Table 1 and the resulting light curve is shown in Figure 1. 

Adopting an approximate ``R-band'' magnitude of 17.1 for maximum light yields a 
red absolute magnitude of $\approx$ $-10.5$  at a 
 distance of $\approx$ 3.2 Mpc for NGC 4214 with A$_{R} = 0.06$ mag  foreground Galactic reddening from the {\it NASA/IPAC Extragalactic Database}\footnote{http://nedwww.ipac.caltech.edu/}.  SN 2010U was clearly sub-luminous for a supernova. Its
maximum luminosity was more like the classical
LBVs in their high mass loss or optically thick wind stage \citep{HD94}. 
However SN 2010U's very rapid decline is not typical of LBVs. 

We located the position of the transient on a deep, archival
pre-outburst {\it HST/WFC3} F814W image of NGC 4214 (Proposal 11360; PI
R. O'Connell) obtained on 2009, Dec. 23, and shown in Figure 2. To solve the
relative astrometry between the post-outburst LBT R-band image
obtained on 2010, Mar. 18, and the pre-outburst WFC3 image, we used 6
bright, isolated point sources around the position of the transient
identified in the WFC3 image. The astrometric solution was obtained using a 2nd
order polynomial transformation with standard tasks in IRAF ({\it geotran} 
and {\it geomap}). The final rms of the solution is 0\farcs013 and 0\farcs021 in the
x and y axis of the WFC3 image, respectively. As can be seen in Figure
2, there is no progenitor identified in the pre-outburst WFC3 image
within the uncertainties of the relative astrometry. We obtained
consistent results using shallower archival WFPC2 and LBT pre-outburst images.

The associated stellar population is shown on the CMD in Figure 3.  
The {\it HST/WFPC2} $F555W$ and $F814W$ archival observations  obtained on 1997 December 9  
 were reduced using the same procedure and with
the same quality cuts as applied to the ANGST sample (details in
\citet{Dalcanton2009}).  The observations  were calibrated and flat-fielded 
using the standard \emph{HST} pipeline and the magnitudes were measured  
using HSTphot \citep{Dolphin2000}.  Only
the highest quality photometry is used for the CMD. The color-coding 
corresponds to the spatial distribution of the stars relative to the 
position of SN 2010U in the lower panel.

We find no visible star at the position of SN 2010U  down to the 50\% completeness
limiting magnitudes of 24.4 and 23.5 of the $F555W$ and $F814W$ images, 
respectively, and thus no candidate precursor. At the distance of NGC 4214,
these magnitude limits imply an upper limit  of $\sim$ $-3.2$ mag to the 
absolute visual magnitude of a possible progenitor. 
Most of the stars within $\approx$ 100 pc
of SN2010U are relatively faint and red, with $F555W-F814W$ colors of $\sim$ 1 -- 2 mag,
suggesting that the likely precursor is associated with an evolved population. 
Of course, it is possible that the progenitor star may have been heavily obscured 
by circumstellar dust prior to its eruption similar to the NGC 300 OT \citep{JP08} and
SN 2008S \citep{JPetal08}, and therefore be an intrinsically more luminous star. However, 
the spatial distribution (Figure 3) of the younger and
more luminous main sequence stars, shows that the precursor star is not 
closely associated with a more massive young population. 
This, however, does not rule out an obscured, evolved star (AGB) of lower mass.
Based on the CMD's, the characteristics of the associated stellar population, 
and the limiting magnitudes of the images,  
we suggest that the progenitor (or its companion) was most likely an evolved intermediate 
mass star, with an upper mass limit of  3 to  5 M$\odot$.  

This conclusion is supported by the post-maximum spectrum obtained on 20 February, 2010, fourteen days after discovery. The  CCDS spectrograph on the 
Hiltner 2.4m MDM telescope was used with the 
150{\it l} grating giving a resolution of $\sim$ 14{\AA} with a 2{\arcsec} slit.
The low resolution spectrum is shown in Figure 4. The spectrum is dominated by
strong emission lines of H$\alpha$ and H$\beta$ plus the O I blend at 
$\lambda$7774{\AA}.  This is not the spectrum of an LBV at maximum nor does
it resemble the spectra of the less luminous transients like SN 2008S and
the NGC 300 - OT. The strong O I emission indicates it is most like a post-maximum nova. The lines are resolved with a FWHM of $\sim$ 1900 km s$^{-1}$, and the 
radial velocities of their line centers are 200 -- 300 km s$^{-1}$. The H$\alpha$ and 
 H$\beta$ line profiles are asymmetric to the red, and  H$\alpha$ has asymmetric wings extending to $-$2300 km s$^{-1}$ and +3200 km s$^{-1}$. The asymmetry in the line profiles and the wings may be due to remnant P Cyg absorption. 
  H$\alpha$ and O I $\lambda$7774{\AA} also have split 
 profiles possibly indicative of a bi-polar outflow or planar disk \citep{Lynch06}. Similar  profiles are often observed in classical novae \citep{Lynch06,Austin96,Della02}. 
 The blue and red emission peaks in the O I line 
are shifted by $-$610 and +460 km s$^{-1}$, respectively, relative to the line 
center. The corresponding shifts in H$\alpha$ are less, $-374$ km s$^{-1}$ and
+150 km s$^{-1}$. Other weaker lines in the spectrum are identified with 
Fe II, [Fe II] and [O I]. The nebular [O III] lines may also be present, but
they are blended with other lines.

\section{Discussion}

The rapid decline of the light curve, the post-maximum spectrum, and 
the relatively low mass and luminosity inferred from the lack of a 
precursor on deep 
pre-eruption images and the associated stellar population, all suggest that
SN 2010U was a luminous nova.  

SN 2010U shares characteristics with the fast novae. If we adopt the  
 observations  by  the amateur observers and assume that it was at or near maximum, then 
 the time to 
decline by two magnitudes (t$_{2}$) is 15 days and the time to decline three 
magnitudes (t$_{3}$) is 26 days. These decline rates are consistent with a  fast 
nova. Classical novae obey the Maximum-Magnitude-Rate of Decline relation (MMRD) 
of \cite{Della95}. Adopting M$_{r}$, of $-10.5$  for maximum light and a $V-R$ color of 1.1 to 1.5 mag from V1500 Cyg, a classical nova a few days past maximum 
light \citep{Jay76}, M$_{v}$ is $\approx$ $-9$  to $-9.4$~mag, near the upper luminosities 
observed for CNe. A t$_{2}$ of 15 days implies a maximum M$_{v}$ of 
$-8.7$ mag \citep{DD2000}, comparable to  SN 2010U's maximum luminosity. 
Of course, it is possible that maximum was missed and SN 2010U was  more 
luminous, but the P Cyg
profiles reported in CBET 2163 indicate that it was not missed by more than
a few days. Furthermore, a typical fast to moderately fast nova, with 
t$_{2}$ $\sim$ 15 days, has emission lines with FWHM from  2000 km s$^{-1}$ 
up to as high as 6000 km s$^{-1}$. SN 2010U is on the low end but 
consistent with  this range. \citet{Kas2010} have recently described a 
group of fast and luminous novae that are apparently inconsistent with the 
MMRD relation. 
However, SN 2010U is at the upper end of luminosities expected
from the MMRD relation for normal novae and was super-Eddington at maximum.

While not a high mass eruptive variable like most of the other ``SN impostors'', 
SN 2010U was indeed an impostor and is an excellent example of the need for 
post--maximum photometry and spectra, and a study of the stellar environment 
to determine the nature and evolutionary state of these objects.

\acknowledgments
We are especially grateful to the observers  Syuichi Nakano, Koichi Itagaki, 
Ken-ichi Kadota, and Toru Yusa for forwarding their discovery images of SN 2010U. 
C. S. Kochanek and K. Z. Stanek  acknowledge  support by NSF grant AST-0908816; Stanek is also supported by NSF grant AST-0707982. J. L. Prieto  acknowledges 
support from NASA through Hubble
Fellowship grant HF-51261.01-A awarded by the STScI, which is operated by
AURA, Inc. for NASA, under contract
NAS 5-26555.

{\it Facilities:}  \facility{LBT, MDM, HST/WFPC2, HST/WFC3}



\begin{deluxetable}{lllll}
\tablecolumns{5}
\tablecaption{Journal of Observations and Magnitudes}
\tablehead{
\colhead{JD}  &  \colhead{Calendar Date}  &  \colhead{Magnitude}   
& \colhead{Filter} & \colhead{Observer or Telescope}}
\startdata
2455233.1    &  2010-02-05 &  17.2  & unfiltered\tablenotemark{a}    &   Itagaki\\   
2455233.2    &  2010-02-05 &  16.9   & unfiltered\tablenotemark{a}    &  Kadota \\  
2455233.2    &  2010-02-05 &  17.2   & unfiltered\tablenotemark{a}    &   Itagaki\\
2455233.8    &  2010-02-06 &  17.1   & unfiltered\tablenotemark{a}   &   Yusa \\  
2455234.1    &  2010-02-06 &  17.3   & unfiltered\tablenotemark{a}  &   Itagaki\\  
2455234.2    &  2010-02-06 &  16.9    & unfiltered\tablenotemark{a} &   Kadota\\  
2455235.9    &  2010-02-08 &  18.53$\pm$ 0.05  &  SDSS-g & MDM\\  
2455235.9    &  2010-02-08 &  17.88$\pm$ 0.02 &  SDSS-r & MDM\\
2455247.9    &  2010-02-20 &  19.64$\pm$ 0.03  & SDSS-r  & MDM\\
2455252.8    &  2010-02-25 &  19.92$\pm$ 0.11  &  SDSS-r  &  MDM\\
2455252.8    &  2010-02-25 &  20.63$\pm$ 0.09  &  SDSS-i  &  MDM\\ 
2455273.9    &  2010-03-18 &  22.59$\pm$ 0.17  & Bessel-U &  LBT\\ 
2455273.9    &  2010-03-18 &  23.20$\pm$ 0.09  & Bessel-B &  LBT \\
2455273.9    &  2010-03-18 &   22.53$\pm$ 0.05 &  Bessel-V &  LBT\\
2455273.9    &  2010-03-18 &   21.35$\pm$ 0.03 &  Bessel R  & LBT\\
\enddata
\tablenotetext{a}{Approximate ``R'' band, see text}
\end{deluxetable} 

\clearpage
\begin{figure}
\plotone{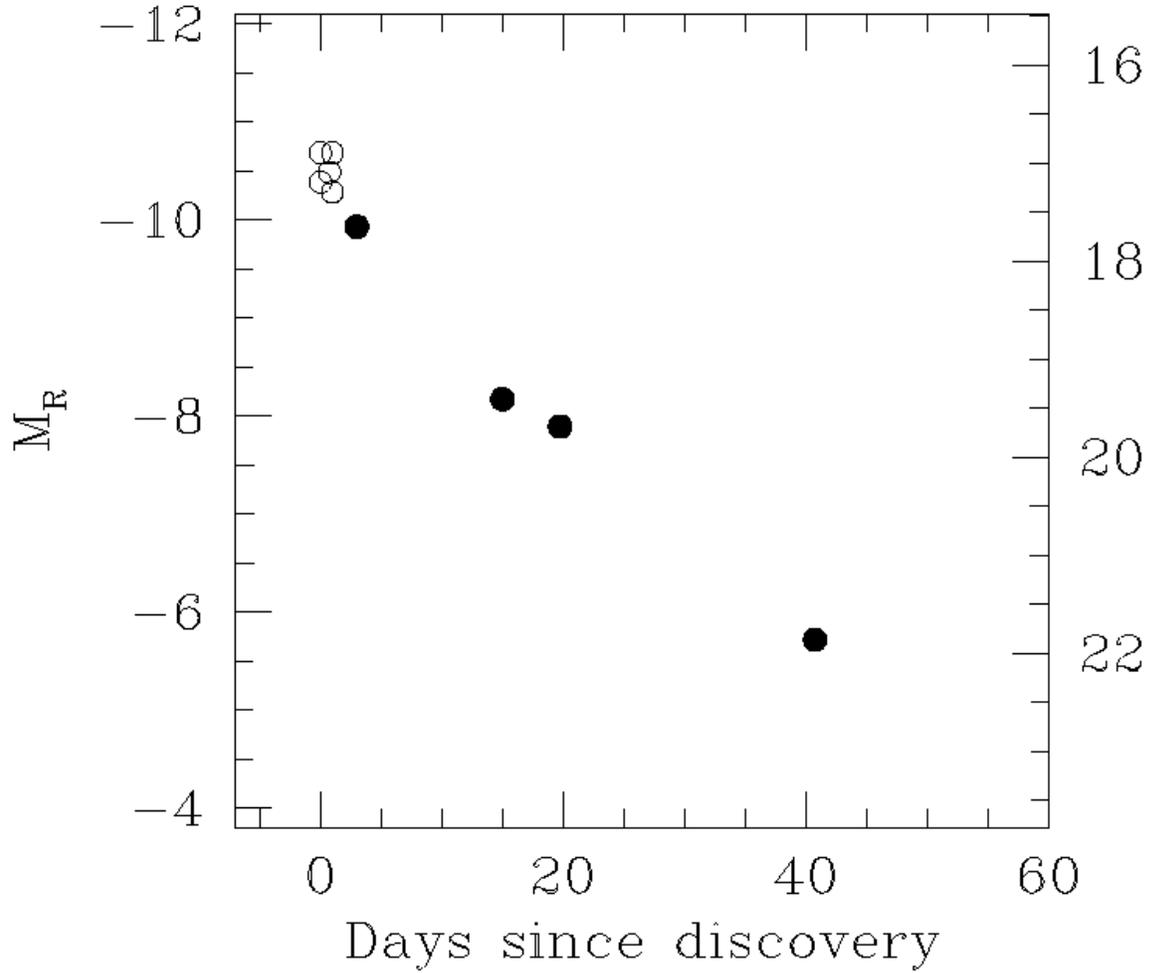}
\caption{The light curve for SN 2010U showing the discovery photometry from CBET 2161 transformed to an approximate ``R'' band (see text) and our CCD red photometry. The MDM r-band magnitudes have been converted to the Bessel R-band.}
\end{figure}

\clearpage
\begin{figure}
\plotone{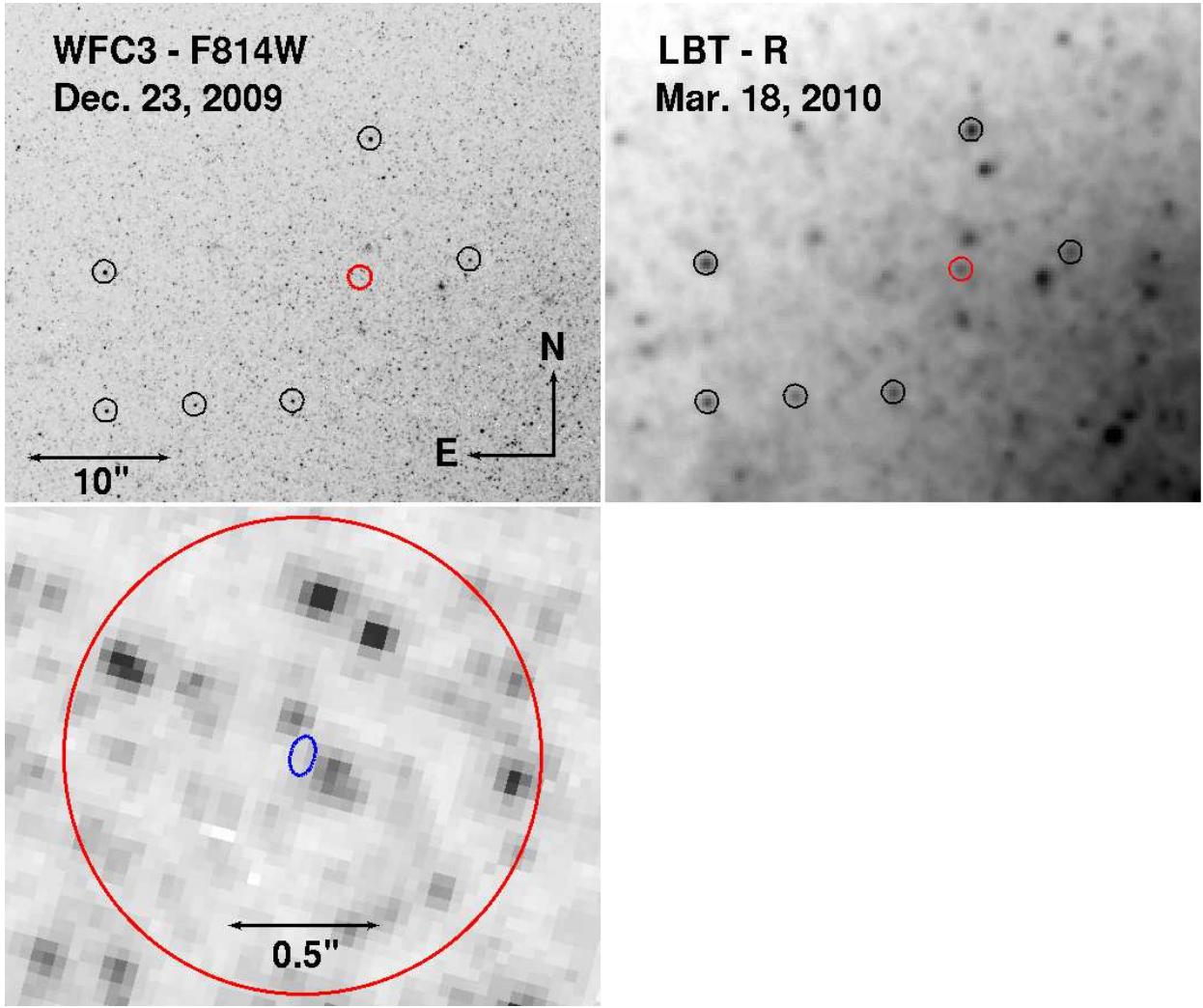}
\caption{Top left: WFC3 I-band (F814W) image from 2009 Dec 23.
The black  circles mark the  six stars used for the relative astrometry. 
The red circle marks the position of the transient
after solving the relative astrometry between WFC3 and LBT
image. All the circles have radius = 0.8". Top right: LBT image from 2010 March 
18 showing SN 2010U and  used to determine the relative astrometry.
The same stars are marked here (circles have the same
radius) and the red circle marks the position of the transient.
Lower left: Zoomed-in WFC3 F814W image around the position of
the transient. The  minor and major axes of the blue ellipse are 
3 times the rms of the astrometric solution. The minor axis radius
is $0\farcs04$, and the major axis radius is $0\farcs06$. The rms
of the astrometric solution is $0\farcs013$ in the x-axis and $0\farcs021$ 
in the y-axis.}
\end{figure}

\begin{figure}
\caption{Left Top: CMD from the archival data field (grey), the colors correspond
to the spatial distribution in the lower panel.  Bottom: Spatial distribution of the stars
in the  archival field (grey) with stars near SN 2010U color coded
by projected distance;   within 50 pc (purple), 100 pc (blue), 200 pc
(green), 300 pc (yellow), 400 pc (orange), 500 pc (red). The location of
SN 2010U is marked with a red star.
Right Top: The CMD with the young main sequence stars; $m_{F814W} <$ 22 (blue) and  22 $< m_{F814W} <$ 23 (green).
Bottom: The spatial distribution of the stars in the above panel with the same
color scheme. }
\end{figure}

\clearpage
\begin{figure}
\epsscale{0.95}
\plotone{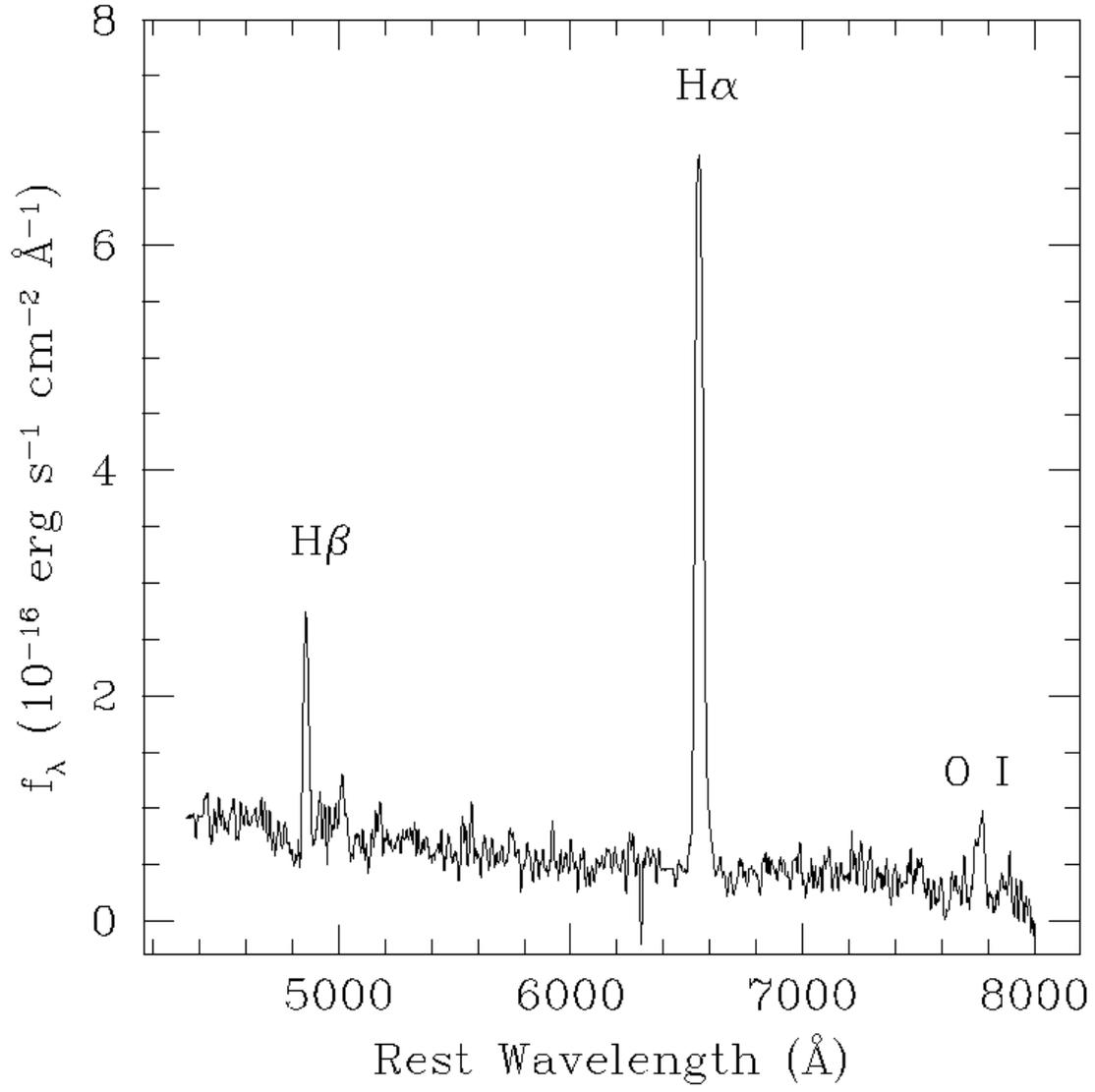}
\caption{The flux calibrated post-maximum spectrum of SN 2010U.}  
\end{figure}

\end{document}